\documentclass[aps,showpacs,showkeys,twocolumn]{revtex4}
\usepackage{graphicx}
\usepackage{amssymb}
\usepackage{bm}

\begin{document}
\setcounter{page}{1}
\title[]{Cosmological Aspects of the D-brane World}
\author{Inyong \surname{Cho}}
\email{iycho@skku.edu}
\author{Yoonbai \surname{Kim}}
\email{yoonbai@skku.edu}
\affiliation{BK21 Physics Research Division and Institute of Basic Science,
Sungkyunkwan University, Suwon 440-746}
\author{Eung Jin \surname{Chun}}
\email{ejchun@kias.re.kr}
\affiliation{Korea Institute for Advanced Study,
207-43 Cheongryangri 2-dong Dongdaemun-gu, Seoul 130-722}
\author{Hang Bae \surname{Kim}}
\email{hbkim@hanyang.ac.kr}
\affiliation{BK21 Division of Advanced Research and Education in Physics,
Hanyang University, Seoul 133-791}
\date[]{Received July 5 2006}

\begin{abstract}
The D-brane world is an idea that we are living on the D-brane
imbedded in a 10- or 11-dimensional spacetime of string theories,
aiming at the construction of realistic models from string theories.
We investigate the cosmological aspects of the D-brane world,
focusing on homogeneous anisotropic cosmology driven
by the dilaton and the NS-NS 2-form field which becomes massive
in the presence of the D-brane.
The dilaton possesses the potential due to the presence of the D-brane,
various form field fluxes, and the curvature of extra dimensions.
In the absence of stabilizing potential, we found the attractor solutions
for this system which show the overall features of general solutions.
In the presence of the non-vanishing NS-NS 2-form field, the homogeneous
universe expands anisotropically while the D-brane term dominates.
The isotropy is recovered as the dilaton rolls down
and the curvature term dominates.
With the stabilizing potential for the dilaton, the anisotropy developed
by the initial NS-NS 2-form field flux is erased as the NS-NS 2-form field
begins to oscillate around the minimum, forming the B-matter,
and the isotropic matter-dominated universe is obtained.
\end{abstract}

\pacs{11.15.Uv, 04.60.-m}

\keywords{D-brane, Antisymmetric tensor field, Dilaton, Cosmology}

\maketitle

\section{Introduction}

The idea of the brane world first appeared as an alternative solution
to the gauge hierarchy problem by allowing large or warped extra
dimensions while we are leaving on a 3-brane to avoid various known
experimental bounds \cite{brane-world}.
String theories have such structures as D-branes on which open strings
can reside and thus provide a framework where realistic brane world models
can be constructed.
The D-brane world is an idea that we are living on the D-brane imbedded
in a 10- or 11-dimensional spacetime of string theories, aiming at
the construction of realistic models from string theory.
One way to see how realistic the D-brane world models can be is to
look into the cosmological aspects of them.

Low energy effective theories derived from the NS-NS sector of
string theories contain the gravity, $g_{\mu\nu}$, the dilation,
$\Phi$, and the NS-NS 2-form field, $B_{\mu\nu}$.
The existence of the last degree of freedom leads to  intriguing
implications in string cosmology \cite{Goldwirth:1993ha}.
In four spacetime dimensions,
the massless antisymmetric 2nd-rank tensor field is dual to the a
pseudo-scalar (axion) field \cite{Kalb:1974yc},
and the axion--dilaton system is known to develop an unobserved anisotropy
in our Universe, which can be diluted away at late times only in a
contracting universe \cite{Copeland:1994km}.

Such a disastrous cosmological situation can be resolved also in
string theory which contains another dynamical object called D-brane
\cite{Witten:1995im}. The gauge invariance on the D-brane is
maintained through the coupling of the gauge field strength to the
NS-NS 2-form field \cite{Cremmer:1973mg,Witten:1995im,Pol}.
Then, the effective action derived on the D-brane describes the NS-NS
2-form field as a massive and self-interacting antisymmetric tensor field.
Cosmological evolution of such a tensor field has been investigated
in Ref.~\cite{Chun:2005ee} assuming that the dilaton is fixed
to a reasonable value. Although the time-dependent magnetic $B$
field existing in the early universe develops an anisotropy in the
universe, it was realized that the matter-like behavior of the $B$
field (B-matter) ensures a dilution of the anisotropy at late times
and thus the isotropy is recovered in reasonable cosmological
scenarios \cite{Chun:2005ee}. In such sense the effect of antisymmetric
tensor field on the D-brane is distinguished from that of field strength
of the U(1) gauge field \cite{Kim:2004zq}.

In this paper, we investigate the cosmological evolution of the
dilaton--NS-NS 2-form field system in our Universe
which is assumed to be imbedded in the D-brane.
In general, the D-brane and the NS-NS 2-form field couple to the R-R
form fields. Here, we assume a trivial R-R background and omit
them in our analysis.
The usual string cosmology with the dilaton suffers from  the notorious
runaway problem, which is also troublesome in the D-brane universe.
In our study, the dilaton obtains two exponential potential terms
due to the curvature of extra dimensions $\Lambda$,
and the D-brane tension (the mass term of the B-matter) $m_B$.
It is interesting to observe that the dilaton can be stabilized
for negative $\Lambda$ \cite{Kim:2005kr} which, however,
leads to a contracting universe due to the effective negative
cosmological constant in our Universe.
When $\Lambda$ is positive, the B-matter dominance
will be overturned by the $\Lambda$ dominance as the dilaton runs
away to the negative infinity. As a consequence of this,
the initial anisotropy driven by the B-matter can also be diluted away
at late times.

For a realistic low energy effective theory,  string theory must
be endowed with a certain mechanism generating an appropriate
vacuum expectation value for the dilaton.  In such a situation,
the dilaton is expected to be stabilized at some stage of the cosmological
evolution affecting the dynamics of the NS-NS 2-form field.
Taking an example of the dilaton stabilization, we will also examine
the cosmological evolution of the dilaton--NS-NS 2-form field system
in which the essential features of Ref.~\cite{Chun:2005ee} are reproduced.

This paper is organized as follows.
In Section 2, we describe the low energy effective action of the D-brane
world and the corresponding field equations.
In Section 3, we find analytic and numerical cosmological solutions
of the dilaton--NS-NS 2-form field system
to observe an intriguing interplay of the curvature $\Lambda$ and the
D-brane tension $m_B$. In Section 4, we consider the evolution of the
anisotropic universe with the dilaton stabilization which leads to
a satisfactory cosmology of the D-brane universe.
We conclude in Section 5.

\section{Effective Field Theory of the D-Brane World}

The main idea of the D-brane world is that we reside on a D$p$-brane
imbedded in 10- or 11-dimensional spacetime with extra-dimensions
compactified. The bosonic NS-NS sector of the D-brane world consists of
the U(1) gauge field $A_\mu$ living on the D$p$-brane and the bulk
degrees including the graviton $g_{\mu\nu}$, the dilaton $\Phi$, and
the antisymmetric tensor field of rank-two $B_{\mu\nu}$.

The low energy effective action consists of the bulk action
\begin{equation}
S_{\rm NS} = \frac{1}{2\kappa_{10}^2} \int d^{10}x \, \sqrt{-g}\, e^{-2\Phi}
\left[R+4(\partial\Phi)^2-\frac{1}{12}H^{2}\right],
\label{act10}
\end{equation}
and the brane action in $p$ spatial dimensions;
\begin{eqnarray}
S_{{\rm D}p} &=& -\mu_p\int d^{p+1}x\, e^{-\Phi}\sqrt{-\det(g+{\cal B})}
\nonumber\\&&{}
+i\mu_p\int_{p+1}{\rm Tr}\left[\exp({\cal B})\wedge\sum_q C_q\right]\,,
\label{dpac}
\end{eqnarray}
where $\kappa_{10}^2={1\over2} (2\pi)^7 \alpha'^4$ and $\mu_p^2=
(\pi/\kappa_{10}^2) (4\pi^2\alpha')^{3-p}$ \cite{Pol}. Recall that
$\alpha'$ defines the string scale; $m_{{\rm s}} =
\alpha'^{-1/2}$. Here the spacetime indices  are not explicitly
expressed and the field strength $H$ of the NS-NS form field
is $H_{\mu\nu\rho}=\partial_{[\mu}{\cal B}_{\nu\rho]}$
because of Bianchi identity of the U(1) gauge field.
In the presence of the brane, the gauge invariance of $B_{\mu\nu}$ is restored
through its coupling to a U(1) gauge field $A_\mu$ and the gauge
invariant field strength is \cite{Witten:1995im}
\begin{equation}
{\cal B}_{\mu\nu} \equiv B_{\mu\nu}+2\pi\alpha'F_{\mu\nu},
{\rm\ where\ } F_{\mu\nu}=\partial_\mu A_\nu-\partial_\nu
A_\mu.
\end{equation}

The D-brane and the NS-NS form field on it couple to the R-R form field
as shown in the second term of Eq.~(\ref{dpac}).
In this paper, we consider the case that the D-brane and the NS-NS form
field are homogeneous in our 4D world and all R-R form fields live in extra
dimensions so that their sole effect is the curvature term in the 4D
effective action which will be discussed below.
Thus all R-R form fields are not considered from now on,
assuming the trivial R-R background.
We leave the more general case of both the NS-NS and the R-R form
field living in our 4D world as our future work.

After the compactification of extra dimensions,
the four-dimensional effective action of the bosonic sector
in the string frame is \cite{Chun:2005ee}
\begin{eqnarray}
S_{{\rm S}}&=& \frac{1}{2\kappa_4^2} \int d^4\tilde
x \sqrt{-\tilde g} \left[ e^{-2\Phi}
\vphantom{\sqrt{\frac12}}\right.\nonumber\\&&\hspace{-5mm}\times
\left(\tilde
R-2\Lambda+4\tilde\nabla_\mu\Phi\tilde\nabla^\mu\Phi -\frac{1}{12}
\tilde H_{\mu\nu\rho}\tilde H^{\mu\nu\rho}\right)
\nonumber\\&&\hspace{-5mm}\left.{}
-{m_B^2}e^{-\Phi} \sqrt{1 + \frac12\tilde{\cal
B}_{\mu\nu}\tilde{\cal B}^{\mu\nu}
 - \frac{1}{16}\left(\tilde{\cal B}^*_{\mu\nu}\tilde{\cal B}^{\mu\nu}\right)^2}\ \right],
\label{action-S}
\end{eqnarray}
where the tilde denotes the string frame quantity
and ${\cal B}^{\ast}_{\mu\nu}=\frac12\sqrt{-g}\epsilon_{\mu\nu\alpha\beta}
{\cal B}^{\alpha\beta}$ with $\epsilon_{0123} = 1$.
The parameter $m_B$ is defined by $m_B^2=2\kappa_4^2\mu_p$.
If we assume that the six
extra-dimensions are compactified with a common radius $R_{{\rm c}}$,
one finds $m_B=\pi^{\frac14}(g_{{\rm s}}^2/4\pi)^{\frac{p-3}{16}}
\left(R_{{\rm c}}M_{{\rm P}}\right)^{\frac{15-p}{8}}M_{{\rm P}}$
where $g_{\rm s}=e^\Phi$ and $M_{{\rm P}}=2.4\times10^{18}{\rm GeV}$ is
the four-dimensional Planck mass.
The qualitative features of our results do not depend on specific
values of $p$.  Thus, we take $p=3$ for simplicity.
The $\Lambda$ term comes from the scalar curvature
or the condensates of the NS-NS form field $\langle H^2 \rangle$
or R-R form fields of extra dimensions integrated over the whole extra dimensions
\begin{equation}
-\frac{\Lambda}{\kappa_4^2} \sim \int dx^6\,
\left[R^{(6)}-\frac{1}{12}\langle H^2\rangle+\cdots\right].
\end{equation}
In the following we treat $\Lambda$ as a free parameter which can be
either positive or negative.

The action, and thus also field equations, can be written in a more
familiar form in the Einstein metric, which is defined by
\begin{equation}
g_{\mu\nu} = e^{-2\Phi}\tilde g_{\mu\nu}.
\end{equation}
We will work in this metric from now on. In terms of Einstein
metric, the action becomes
\begin{eqnarray}
\label{action-E}
S_{\rm E}&=& \frac{1}{2\kappa_4^2} \int d^4x
\sqrt{-g} \left[ R -2(\nabla\Phi)^2 -\frac{1}{12}e^{-4\Phi}H^2
\vphantom{\sqrt{\frac12}}\right.\hspace{10mm}\nonumber\\&&\hspace{-10mm}\left.
-2\Lambda e^{2\Phi}
-{m_B^2}e^{3\Phi}\sqrt{1+\frac12e^{-4\Phi}{\cal B}^2
- \frac{1}{16}e^{-8\Phi}\left({\cal B}^*{\cal B}\right)^2} \right].
\end{eqnarray}
The field equations derived from the action (\ref{action-E}) are
\begin{eqnarray}
&&\hspace{-8mm}
\nabla^\lambda H_{\lambda\mu\nu}
-4H_{\lambda\mu\nu}\nabla^\lambda\Phi
\nonumber\\&&
-m_B^2e^{3\Phi}\frac{{\cal B}_{\mu\nu}-\frac14e^{-4\Phi}{\cal B}^*_{\mu\nu}
\left({\cal B}{\cal B}^*\right)}{\sqrt{1+\frac12e^{-4\Phi}{\cal B}^2
-\frac{1}{16}e^{-8\Phi}\left({\cal B}{\cal B}^*\right)^2}} = 0,
\label{B-eq1}
\end{eqnarray}
\begin{equation}
\label{P-eq1}
-\nabla^2\Phi + \frac{\partial V(\Phi)}{\partial\Phi}= 0,
\end{equation}
\begin{equation}
\label{E-eq1}
G_{\mu\nu} = \kappa_4^2 T_{\mu\nu},
\end{equation}
where the dilaton potential is
\begin{eqnarray}
V(\Phi) &=& \frac14\left[2\Lambda e^{2\Phi} +
\frac{1}{12}e^{-4\Phi}H^2
\vphantom{\sqrt{\frac12}}\right.\nonumber\\&&\hspace{-15mm}\left.{}
+m_B^2e^{3\Phi}\sqrt{1 + \frac12e^{-4\Phi}{\cal B}^2 -
\frac{1}{16}e^{-8\Phi}\left({\cal B}^*{\cal B}\right)^2} \ \right],
\label{dilaton-potential}
\end{eqnarray}
and the energy-momentum tensor is given by
\begin{eqnarray}
\kappa_4^2 T_{\mu\nu}&=& -g_{\mu\nu}\Lambda e^{2\Phi}
+2\nabla_\mu\Phi\nabla_\nu\Phi-g_{\mu\nu}(\nabla\Phi)^2
\hspace{10mm}\nonumber\\ && {}
+\frac{1}{12}e^{-4\Phi}
\left(3H_{\mu\lambda\rho}H_\nu^{\;\lambda\rho}
-\frac12g_{\mu\nu}H^2\right)
\nonumber\\ && \hspace{-15mm} {}
+\frac12m_B^2e^{3\Phi}
\frac{-g_{\mu\nu}-\frac12g_{\mu\nu}e^{-4\Phi}{\cal B}^2
+e^{-8\Phi}{\cal B}_{\mu\lambda}{\cal B}_\nu^{\;\lambda}}
{\sqrt{1+\frac12e^{-4\Phi}{\cal B}^2
-\frac{1}{16}e^{-8\Phi}\left({\cal B}{\cal B}^*\right)^2}}.
\label{TB}
\end{eqnarray}

In the subsequent sections, we examine the equations of motion
(\ref{B-eq1})--(\ref{E-eq1}) and find cosmological homogeneous solutions.
Cosmological implications of the D-brane world is of our main
interest, including the stabilization of the dilaton and the evolution
of the 2-form field.

\section{Cosmological Homogeneous Solutions}

\subsection{The equations of motion}

In this section we study cosmological solutions of the D-brane universe
in the presence of both the dilaton and the NS-NS form field.
We assume spatially homogeneous configurations
for the NS-NS form field and the dilaton, and look for the
time evolution of these fields and the expansion of the universe.

The non-vanishing homogeneous antisymmetric tensor field, in
general, implies the anisotropic universe. To consider the simplest
form of anisotropic cosmology, we take only a single magnetic component of
${\cal B}_{\mu\nu}$ to be nonzero, namely ${\cal B}_{12}(t)\equiv B(t)$
and ${\cal B}_{0i}(t)={\cal B}_{23}(t)={\cal B}_{31}(t)=0$. Then the metric
consistent with this choice of field configuration is of Bianchi
type I
\begin{equation}
ds^2 = -dt^2 + \sum_{i=1}^3 a_i(t)^2(dx^i)^2.
\end{equation}

Then the field equation for $B$, Eq.~(\ref{B-eq1}) becomes
\begin{equation}
\ddot B + \left(-\frac{\dot a_1}{a_1}-\frac{\dot a_2}{a_2}
+\frac{\dot a_3}{a_3}-4\dot\Phi\right)\dot B
+\frac{m_B^2e^{3\Phi}B}{\sqrt{1+B^2/e^{4\Phi}a_1^2a_2^2}} = 0,
\label{eq=Beq}
\end{equation}
the dilaton-field equation (\ref{P-eq1}) is
\begin{equation}
\ddot\Phi+\left(\frac{\dot a_1}{a_1}+\frac{\dot a_2}{a_2}
+\frac{\dot a_3}{a_3}\right)\dot\Phi
= -2\varrho_B-\frac12\varrho_b-\tilde\varrho_b-\varrho_\Lambda,
\label{eq=Phieq}
\end{equation}
and Einstein equations (\ref{E-eq1}) are
\begin{eqnarray}
\frac{\dot a_1}{a_1}\frac{\dot a_2}{a_2}+\frac{\dot a_2}{a_2}\frac{\dot a_3}{a_3}
+\frac{\dot a_3}{a_3}\frac{\dot a_1}{a_1}
&=& \varrho_\Phi +\varrho_B +\varrho_b + \varrho_\Lambda,
\hspace{10mm}\label{eq=G00}\\
\frac{\ddot a_2}{a_2}+\frac{\ddot a_3}{a_3}+\frac{\dot a_2}{a_2}\frac{\dot a_3}{a_3}
&=& -\varrho_\Phi +\varrho_B +\tilde{\varrho}_b + \varrho_\Lambda,
\label{eq=G11}\\
\frac{\ddot a_3}{a_3}+\frac{\ddot a_1}{a_1}+\frac{\dot a_3}{a_3}\frac{\dot a_1}{a_1}
&=& -\varrho_\Phi +\varrho_B +\tilde{\varrho}_b + \varrho_\Lambda,
\label{eq=G22}\\
\frac{\ddot a_1}{a_1}+\frac{\ddot a_2}{a_2}+\frac{\dot a_1}{a_1}\frac{\dot a_2}{a_2}
&=& -\varrho_\Phi -\varrho_B +\varrho_b + \varrho_\Lambda, \label{eq=G33}
\end{eqnarray}
where
\begin{eqnarray}
&\displaystyle
\varrho_\Lambda = \Lambda e^{2\Phi},\
\varrho_\Phi = \dot\Phi^2,\
\varrho_B = \frac{e^{-4\Phi}\dot B^2}{4a_1^2a_2^2},
&\nonumber\\&\displaystyle
\varrho_b = \frac12m_B^2e^{3\Phi}\left(1+\frac{e^{-4\Phi}B^2}{a_1^2a_2^2}\right)^{1/2},
&\nonumber\\&\displaystyle
\tilde\varrho_b = \frac12m_B^2e^{3\Phi}
\left(1+\frac{e^{-4\Phi}B^2}{a_1^2a_2^2}\right)^{-1/2}.
& \label{br1}
\end{eqnarray}

When $m_B\ne0$,
it is more convenient to employ the dimensionless time variable $\tilde t=m_Bt$
and to introduce the variables $\alpha_i$ and $b$ defined by
\begin{equation}
\alpha_i=\ln a_i,\qquad b=\frac{e^{-2\Phi}B}{a_1a_2}.
\end{equation}
Then the equations of motion are written as
\begin{eqnarray}
\ddot b + (\dot\alpha_1+\dot\alpha_2+\dot\alpha_3)\dot b
\hspace{35mm}&&\hspace{8mm}\nonumber\\{}
+\Bigg[ 2\ddot\Phi+2(-\dot\alpha_1-\dot\alpha_2+\dot\alpha_3-2\dot\Phi)\dot\Phi
\hspace{10mm}&&\nonumber\\{}
+\ddot\alpha_1+\ddot\alpha_2+(\dot\alpha_1+\dot\alpha_2)\dot\alpha_3
+\frac{e^{3\Phi}}{\sqrt{1+b^2}}\Bigg]b &=& 0,
\label{eq-b}
\end{eqnarray}
\begin{equation}
\label{eq-phi}%
\ddot\Phi + (\dot\alpha_1+\dot\alpha_2+\dot\alpha_3)\dot\Phi =
-2\rho_B-\rho_\Lambda-\frac12\rho_b-\tilde\rho_b,
\end{equation}
\begin{eqnarray}
\dot\alpha_1\dot\alpha_2+\dot\alpha_2\dot\alpha_3+\dot\alpha_3\dot\alpha_1
&=& \rho_\Phi+\rho_B+\rho_\Lambda+\rho_b, \label{eq-constraint}\\
\ddot\alpha_1+\dot\alpha_1(\dot\alpha_1+\dot\alpha_2+\dot\alpha_3)
&=& \rho_\Lambda+\rho_b, \\
\ddot\alpha_2+\dot\alpha_2(\dot\alpha_1+\dot\alpha_2+\dot\alpha_3)
&=& \rho_\Lambda+\rho_b, \\
\ddot\alpha_3+\dot\alpha_3(\dot\alpha_1+\dot\alpha_2+\dot\alpha_3)
&=& 2\rho_B+\rho_\Lambda+\tilde\rho_b,
\label{eq-a3}
\end{eqnarray}
where
\begin{eqnarray}
&\displaystyle
\rho_\Lambda=\lambda e^{2\Phi},\quad
\rho_\Phi=\dot\Phi^2,\quad
&\nonumber\\&\displaystyle
\rho_B=\frac14\left[\dot
b+(\dot\alpha_1+\dot\alpha_2+2\dot\Phi)b\right]^2,\quad
&\nonumber\\&\displaystyle
\rho_b=\frac12e^{3\Phi}(1+b^2)^{1/2},\quad
\tilde\rho_b=\frac12e^{3\Phi}(1+b^2)^{-1/2}, &
\end{eqnarray}
where $\lambda=\Lambda/m_B^2$.
In the subsequent subsections, we look for the solutions to the above
field equations for various cases beginning with some simple
solutions.

\subsection{The massless limit ($m_B=0$)}

To see the effect of the brane on the spacetime dynamics, let us
first consider the limit of $m_B=0$ which corresponds to either the
absence of the brane or the limit of vanishing string coupling
$g_{{\rm s}}$, namely the usual massless antisymmetric
tensor field. In this limit, the equation (\ref{eq=Beq}) is easily
integrated to yield a constant of motion
\begin{equation} \label{L3}
\frac{a_3\dot B}{a_1a_2} \equiv L_3 \ (\textrm{constant}).
\end{equation}
With the vanishing potential, $B$ manifests itself by
non-vanishing time derivatives. In the dual variable, it
corresponds to the homogeneous gradient along $x^3$-direction. The
spacetime evolution with the dilaton rolling in this case was
studied in Ref.~\cite{copeland:1995}. Here we have assumed that
the dilaton is stabilized by some mechanism.
Following Ref.~\cite{copeland:1995}, we introduce a new time coordinate
$\eta$ via the relation $d\eta=L_3dt/a_1a_2a_3$. Then the
equations (\ref{eq=G00})--(\ref{eq=G33}) can be written as
\begin{eqnarray}
\label{Eeq123}
\alpha_1'\alpha_2'+\alpha_2'\alpha_3'+\alpha_3'\alpha_1'
&=& \frac14a_1^2a_2^2, \\
\alpha_1''=\alpha_2'' &=& 0, \\
\label{Eeq333}
\alpha_3'' &=& \frac12a_1^2a_2^2,
\end{eqnarray}
where the prime denotes the differentiation with respect to $\eta$.

The solutions for $\alpha_1$ and $\alpha_2$ are trivial
\begin{equation}
\alpha_1 = C_1\eta, \quad
\alpha_2 = C_2\eta,
\end{equation}
where $C_{1,2}$ are constants and we omitted the integration
constants corresponding simply to re-scaling of scale factors. The
$\alpha_3$-equation (\ref{Eeq333}) is also easily integrated to
give
\begin{equation}
\alpha_3 = \frac{e^{2(C_1+C_2)\eta}}{8(C_1+C_2)^2}+C_3\eta .
\end{equation}
The constraint equation (\ref{Eeq123}) relates $C_{1,2}$ and $C_3$ by
$C_3=-C_1C_2/(C_1+C_2)$.
Then the relation between $\eta$ and $L_3t$ is explicitly given by
\begin{eqnarray}
L_3t &=& \int^\eta d\eta\; a_1(\eta)a_2(\eta)a_3(\eta) \nonumber\\
&=& \int^\eta d\eta \exp\left[
(C_1+C_2+C_3)\eta+\frac{e^{2(C_1+C_2)\eta}}{8(C_1+C_2)^2}\right]
\nonumber\\
&=& \left[8(C_1+C_2)\right]^{\frac{C_1+C_2+C_3}{C_1+C_2}}
\int^x dy\; y^{-P}e^y,
\end{eqnarray}
where $x=e^{2(C_1+C_2)\eta}/8(C_1+C_2)^2$ and $P=(C_1+C_2-C_3)/2(C_1+C_2)$.
The evolution of scale factors for large $L_3t$ is given by
\begin{equation}
a_{1,2} \propto \left(\log L_3t\right)^{q_{1,2}}, \qquad
a_3 \propto L_3t,
\end{equation}
where $q_{1,2}=C_{1,2}/2(C_1+C_2)$. Therefore, with non-vanishing
$B_{12}(t)$, only $a_3$ grows significantly and the spatial
anisotropy develops.  This can be seen clearly by considering the
ratio $H_{3}/H_{1,2}$ where $H_i\equiv \dot{a}_i/a_i$.  Taking
$C_1=C_2$ and thus $q_{1,2}=1/4$, we get
\begin{equation}
 {H_{3}\over H_{1,2}}= {4 \log L_3 t}
\end{equation}
which grows as time elapses. Finally, we remark that such
anisotropy cannot be overcome by some other type of the isotropic
energy density in the expanding universe.
Assume that the isotropic universe ($a_i=a$) is driven by, e.g,
radiation energy density $\rho_{{\rm R}}$.
Then, one finds $\rho_B \propto 1/a^2$ from (\ref{br1}) and (\ref{L3})
and thus $\rho_B/\rho_{{\rm R}} \propto a^2$,
which implies that the late-time isotropic solution can be
realized only in a contracting universe \cite{copeland:1995}.

\subsection{The $B$ oscillation}

Let us now take into consideration the effect of space-filling D-brane.
To get sensible solution, we fine-tune the bulk
cosmological constant term to cancel the brane tension, that is
$\Lambda=-m_B^2/2$, so that the effective four-dimensional
cosmological constant vanishes.
We assume again the dilaton is stabilized in some way and set $\alpha_1=\alpha_2$.
Then we can rewrite the full equations as follows
\begin{eqnarray}
\dot\alpha_1^2+2\dot\alpha_1\dot\alpha_3 &=& \frac14\left(\dot b+2\dot\alpha_1b\right)^2
\nonumber\\&&\hspace{0mm}{}
+\frac{1}{2}\left(\sqrt{1+b^2}-1\right),
\label{Eeq-11}\\
\ddot\alpha_1+\dot\alpha_1\left(2\dot\alpha_1+\dot\alpha_3\right) &=&
\frac{1}{2}\left(\sqrt{1+b^2}-1\right),
\label{Eeq-1}\\
\ddot\alpha_3+\dot\alpha_3\left(2\dot\alpha_1+\dot\alpha_3\right)
&=& \frac12\left(\dot b+2\dot\alpha_1b\right)^2
\nonumber\\&&\hspace{0mm}{}
+\frac{1}{2}\left(\frac{1}{\sqrt{1+b^2}}-1\right),
\label{Eeq-3}
\end{eqnarray}
\begin{equation}
\label{Beq-2}
\ddot b+\left(2\dot\alpha_1+\dot\alpha_3\right)\dot b+
\left(2\ddot\alpha_1+2\dot\alpha_1\dot\alpha_3+\frac{1}{\sqrt{1+b^2}}\right)b=0.
\end{equation}

First we examine the evolution of $b(t)$ and scale factors qualitatively.
Suppose $b$ starts to roll from an
initial value $b_0$, while the universe is isotropic in the sense
that $\dot\alpha_{10}=\dot\alpha_{30}$. We assume initially
$a_{10}=a_{30}=1$ ($\alpha_{10}=\alpha_{30}=0$) and $\dot B_0=0$
so that $b_0=B_0$ and $\dot b_0=-2\dot\alpha_{10}B_0$. While $b$
is much larger than unity, the rapid expansion of $\alpha_1$ due
to the large potential proportional to $m_B^2b$ drives $b$ in
feedback to drop very quickly to a small value of order one.
A numerical analysis shows that this happens within $m_Bt<2$
up to reasonably large value of $b_0$ for which the
numerical solution is working. The behavior of $b(t)$ after this
point is almost universal irrespective of the initial value $b_0$
if it is much larger than unity.

Once $b$ becomes smaller than unity, the quadratic term of mass
dominates over the expansion and $b$ begins to oscillate about
$b=0$. Then the expansion of the universe provides the slow
decrease of the oscillation amplitude. The situation is the same
as that of the coherently oscillating scalar field such as the
axion or the moduli in the expanding universe. For small $b$, the
energy-momentum tensor of the oscillating $B$ field is given by
${T_\mu}^\nu={\rm diag}[-\rho,p_1,p_2,p_3]$ where
\begin{eqnarray}
\rho &=& \frac14\left(\dot b+2 \dot\alpha_1b\right)^2
    + \frac12m_B^2\left(\sqrt{1+b^2}-1\right)
\nonumber\\
    &\approx& \frac14\left(\dot b^2+m_B^2b^2\right), \\
p_1=p_2 &=& -\frac14\left(\dot b+2\dot\alpha_1b\right)^2
    - \frac12m_B^2\left(\frac{1}{\sqrt{1+b^2}}-1\right)
\nonumber\\
    &\approx& -\frac14\left(\dot b^2-m_B^2b^2\right), \\
p_3 &=& \frac14\left(\dot b+2\dot\alpha_1b\right)^2
    - \frac12m_B^2\left(\sqrt{1+b^2}-1\right)
\nonumber\\
    &\approx& \frac14\left(\dot b^2-m_B^2b^2\right).
\end{eqnarray}
With the expansion of the universe neglected, the equation of
motion for $b$, Eq.~(\ref{Beq-2}) is then approximated by
\begin{equation}
\label{Beq-osc}
\ddot b+m_B^2b\approx0.
\end{equation}
Since the oscillation is much faster than the expansion, we can
use the time-averaged quantities over one period of oscillation
for the evolution of spacetime. The equation (\ref{Beq-osc}) gives
the relation $\langle\dot b^2\rangle=\langle m_B^2b^2\rangle$.
Thus, the oscillating $B$ field has the property
$p_1,p_2,p_3\approx0$ and behaves like homogeneous and isotropic
matter. This justifies the name of {\em B-matter}. Therefore,
after $b$ begins to oscillate, the isotropy of the universe is
recovered.

\subsection{Attractor solutions}

Now, we consider the generic case that both the dilaton and
the NS-NS 2-form field evolve in time.
The dilaton in the system of equations (\ref{eq-b})--(\ref{eq-a3})
has the exponential potential up to the correction due to the
NS-NS 2-form field. It is well-known that the scalar field
with the exponential potential possesses the scaling solution in
which the energy density of the scalar field mimics the background
fluid energy density \cite{Copeland:1997et,Copeland:2006wr}. This scaling solution is
also an attractor, so that the late time behavior of the
solutions are universal irrespective of initial conditions.
This is an attractive property of the exponential potential. For the
potential of the form $V(\Phi)=V_0e^{\beta\Phi}$, there is an
attractor solution
\begin{equation}
\label{sol-a}
\Phi=-\frac{2}{\beta}\ln\left[\frac{\beta
V_0^{1/2}t}{\sqrt{2(12-\beta^2)}}\right],\quad
\alpha=\left(\frac{2}{\beta}\right)^2\ln t,
\end{equation}
for $0\le\beta<\sqrt{12}\,$. The scale factor obeys the power-law
time-dependence, implying that the rolling of $\Phi$ constitutes
the matter having an equation of state $p=w\rho$ where $w=\beta^{2}/6-1$
varies from $-1$ to $+1$ for the aforementioned range of $\beta$.

We found this type of particular solutions
of the Eqs.~(\ref{eq-b})--(\ref{eq-a3}),
which can be found when we have a single exponential term in the potential,
that is, for the case of $\Lambda=0$ and for the case of $m_B=0$.
For both cases we start from an ansatz of the form
\begin{eqnarray}
&\displaystyle
\alpha_1=\gamma_1\ln\tilde t,\quad
\alpha_3=\gamma_3\ln\tilde t,
&\nonumber\\&\displaystyle
\Phi(\tilde t)=\gamma_\Phi\ln\tilde t+\Phi_0,\quad
b={\rm constant},
\end{eqnarray}
where we suppressed constant terms in $\alpha_1$ and $\alpha_3$
which correspond to simple rescaling of coordinates. For the case of
$\Lambda=0$, we obtain two distinguished solutions
\begin{eqnarray}
&\displaystyle
\alpha_1(\tilde t)=\alpha_3(\tilde t)=\frac{4}{9}\ln\tilde t,
&\nonumber\\&\displaystyle
\Phi(\tilde t)=-\frac{2}{3}\ln\tilde t+\ln\frac{2}{3},\quad
b=0,
\label{sol-1}
\end{eqnarray}
and
\begin{eqnarray}
&\displaystyle
\alpha_1(\tilde t)=\frac{10}{21}\ln\tilde t,\quad%
\alpha_3(\tilde t)=\frac{3}{7}\ln\tilde t,
&\nonumber\\&\displaystyle
\Phi(\tilde t)=-\frac{2}{3}\ln\tilde t
 + \frac{2}{3}\ln\left(\frac{8\cdot5^{1/4}}{21}\right),\quad
b=\pm\frac12.
\label{sol-2}
\end{eqnarray}
The first solution is nothing but the solution (\ref{sol-a}) with $\beta=3$.
The second solution has the non-vanishing NS-NS form field.
For the case of $m_B=0$, we introduce a new rescaled dimensionless time
variable ${\bar t}=\Lambda^{1/2}t$ instead of ${\tilde t}$
and then get the continuous set of solutions
from Eqs.~(\ref{eq-b})--(\ref{eq-a3})
\begin{eqnarray}
&\displaystyle
\alpha_1({\bar t})=\alpha_3({\bar t})=\ln{\bar t},
&\nonumber\\&\displaystyle
\Phi({\bar t})=-\ln{\bar t}+\frac{1}{2}\ln2,\quad
b={\rm arbitrary\ constant}.
\label{sol-3}
\end{eqnarray}
$\Phi({\bar t})$ and $\alpha({\bar t})$ are same as those in
Eq.~(\ref{sol-a}) with $\beta=2$,
while we have the non-vanishing $B$ field condensate.

These solutions are the solutions to the
Eqs.~(\ref{eq-b})--(\ref{eq-a3}) for the specific initial
conditions. However, the importance of these solutions, as noted in
the paragraph above, arises from the fact that they are attractors,
which means that after enough time the solutions with different initial
conditions approach these solutions.
We will confirm this through numerical analysis in the next subsection.

The solution (\ref{sol-1}) applies for the brane tension dominated case
where the dilaton potential is approximated by
$V(\Phi)=\frac12m_B^2e^{3\Phi}$.
The evolution of the dilaton under this potential produces matter with
the equation of state $p=\frac12\rho$.

Once the antisymmetric tensor field is turned on, the anisotropy
appears as in the solution (\ref{sol-2}).
The measure of anisotropy is
\begin{equation}
\frac{\dot\alpha_3}{\dot\alpha_1}=\frac{9}{10}.
\end{equation}
This result is contrasted with that in Ref.~\cite{Chun:2005ee}
where the dilaton is assumed to be stabilized.
The rolling of the dilaton makes the difference.
It affects the dynamics of $B$ field in such
a way that $b$ remains constant at $b=\pm1/2$ instead of oscillating
about the potential minimum $b=0$ and the anisotropy is maintained.

Let us turn to the third solution (\ref{sol-3}). It is relevant when
the dilaton potential arising from the curvature of extra
dimensions, $V(\Phi)=\Lambda e^{2\Phi}$, dominates over other
contributions. In our scheme this happens as the dilaton rolls down
the potential. When $b$ vanishes, the transition point at which
$\Lambda e^{2\Phi}$ starts to dominate over $\frac12m_B^2e^{3\Phi}$ is at
$\Phi_t=\ln(2\lambda)$. Thus, this solution describes the late time
behavior of all the solutions with various initial conditions when
$\Lambda$ is positive.
It is very intriguing since we achieve the isotropic universe in the end.
The anisotropy is the result of the non-vanishing $B$ field,
whose coupling is dictated by the brane tension term.
The rolling of dilaton makes the brane tension term
less important than the extra dimensional curvature term,
which recovers the isotropy.
The rolling of the dilaton under the potential $\Lambda e^{2\Phi}$
now forms the matter having the equation of state $p=-\frac13\rho$,
thus giving marginal inflation.

\subsection{Numerical analysis}

\subsubsection{Initial conditions}

We have the second order differential equations for four variables
$\Phi(t)$, $B(t)$, $\alpha_1(t)=\alpha_2(t)$, $\alpha_3(t)$. Thus we
need eight initial values $\Phi_0$, $\dot\Phi_0$, $B_0$, $\dot B_0$,
$\alpha_{10}$, $\dot\alpha_{10}$, $\alpha_{30}$, $\dot\alpha_{30}$
to specify the solution. Among these, $\alpha_{i0}$ can always be
set to zero by coordinate rescaling. $\dot\alpha_{i0}$ must obey
the constraint equation (\ref{eq-constraint}), but this does not fix
the ratio $\dot\alpha_{10}/\dot\alpha_{30}$.
We choose the isotropic universe as a natural initial condition
which leads to $\dot\alpha_{10}=\dot\alpha_{30}$.

Since the dilaton potential is composed of exponential terms, the shift of
the dilaton field by a constant can be traded for the redefinition of
mass scale. We use this property to take the initial value of the
dilaton to be zero without loss of generality. In our numerical
analysis, we take the dimensionless time variable as
$\tilde t\equiv\tilde mt$ where $\tilde m=m_Be^{\frac32\Phi_0}$
and use the variable $\tilde\Phi\equiv\Phi-\Phi_0$ with its initial value
$\tilde\Phi_0=0$. This means that the proper time scale for the
cosmological evolution is not $m_B^{-1}$, but $\tilde m^{-1}$.
For convenience's sake, we take the initial time as $\tilde t_0=1$.
The other mass scale $\Lambda$ is also affected by this shift
and we can treat it by replacing the parameter $\lambda$ with
$\tilde\lambda\equiv\lambda e^{-\Phi_0}$.

Now we need three initial values $\dot\Phi_0$, $B_0$, and $\dot
B_0$, to fix the functional form of the solution.
The initial values $b_0$ and $\dot b_0$ are related to
$B_0$ and $\dot B_0$ by $b_0=B_0$ and
$\dot b_0=\dot B_0-2(\dot\alpha_{10}+\dot\Phi_0)B_0$.

\subsubsection{Numerical solutions for $B\ne0$}

We turn on the NS-NS 2-form field along the $x^3$ direction,
${\cal B}_{12}=B\ne 0$. The spacetime becomes anisotropic,
$a_1(t)=a_2(t)\ne a_3(t)$, in general.
The solutions are classified by the signature of $\Lambda$.
For $\Lambda<0$, the solution becomes singular.
Here we skip the description of such singular solutions which are not
suitable for the evolution of our Universe.

\begin{figure*}
\begin{center}
\includegraphics[height=70mm]{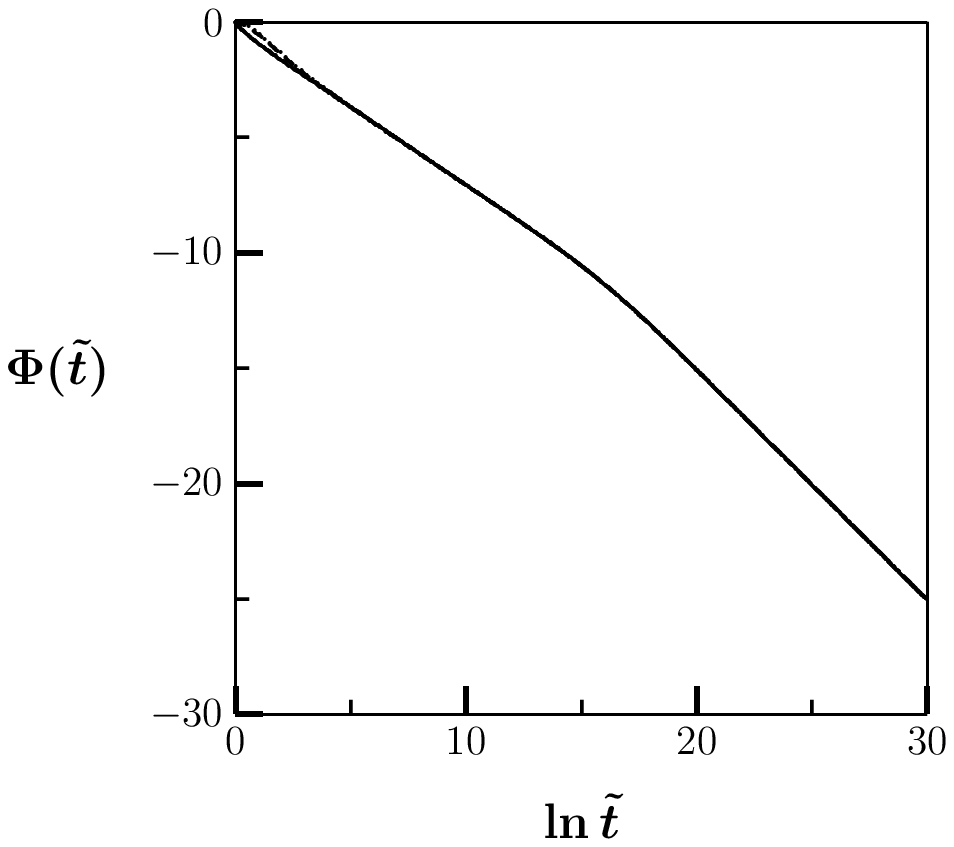}\hspace{10mm}
\includegraphics[height=70mm]{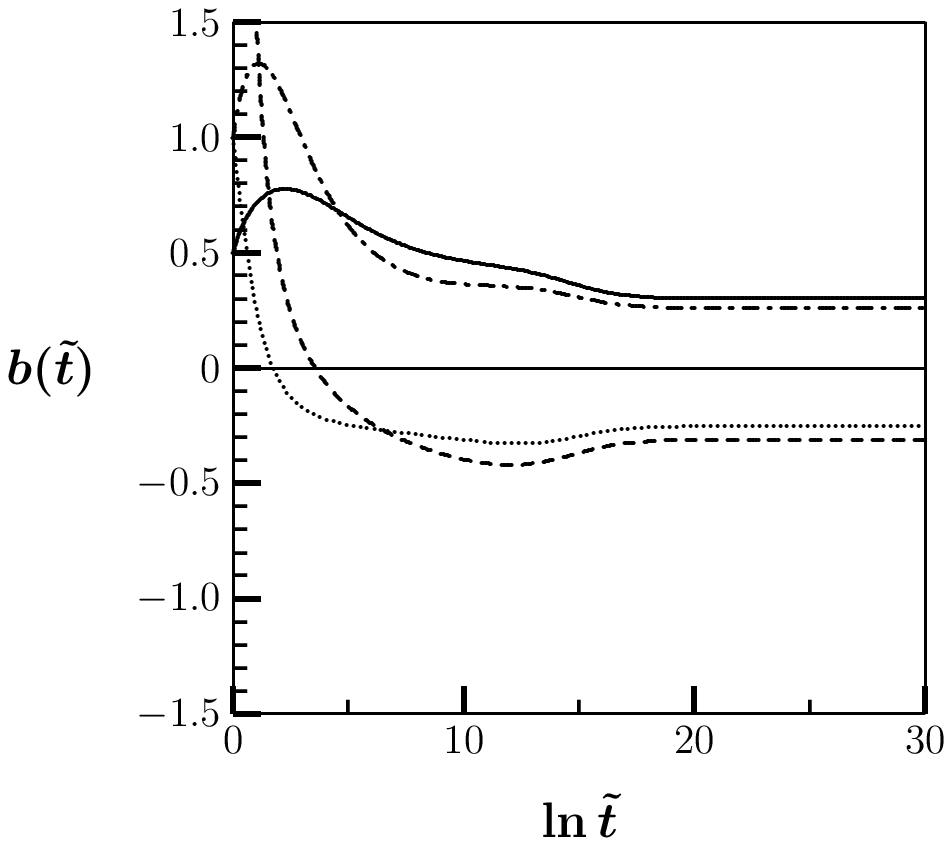}\\[10mm]
\includegraphics[height=70mm]{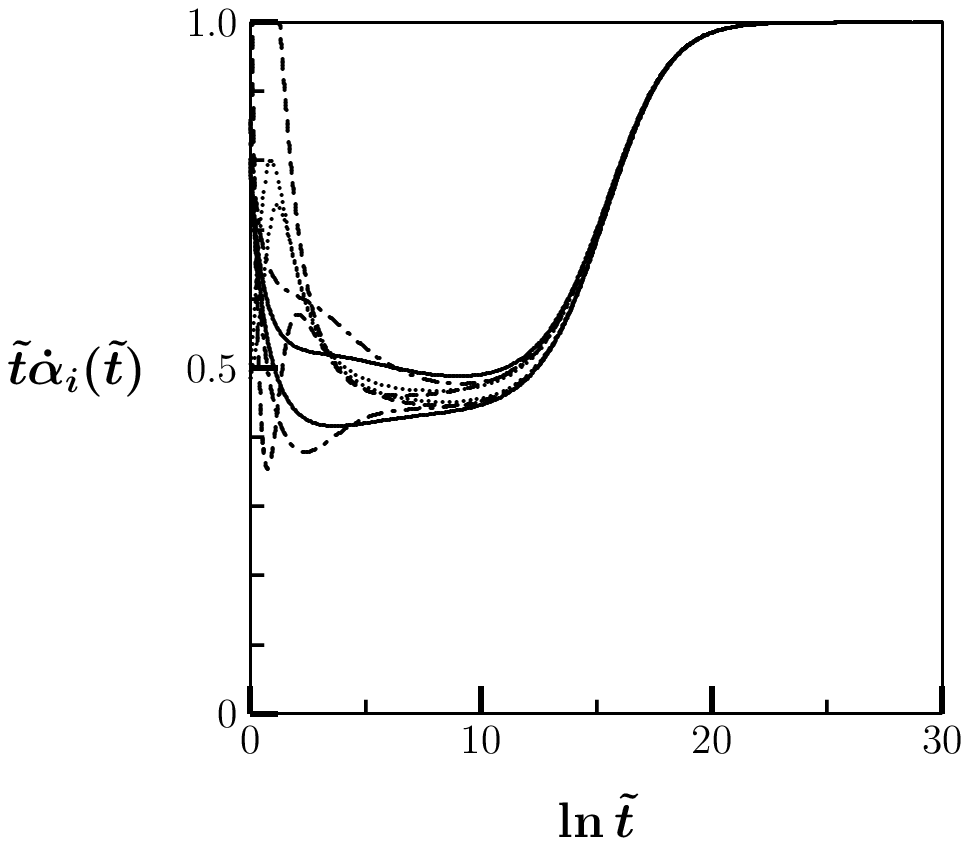}\hspace{10mm}
\includegraphics[height=70mm]{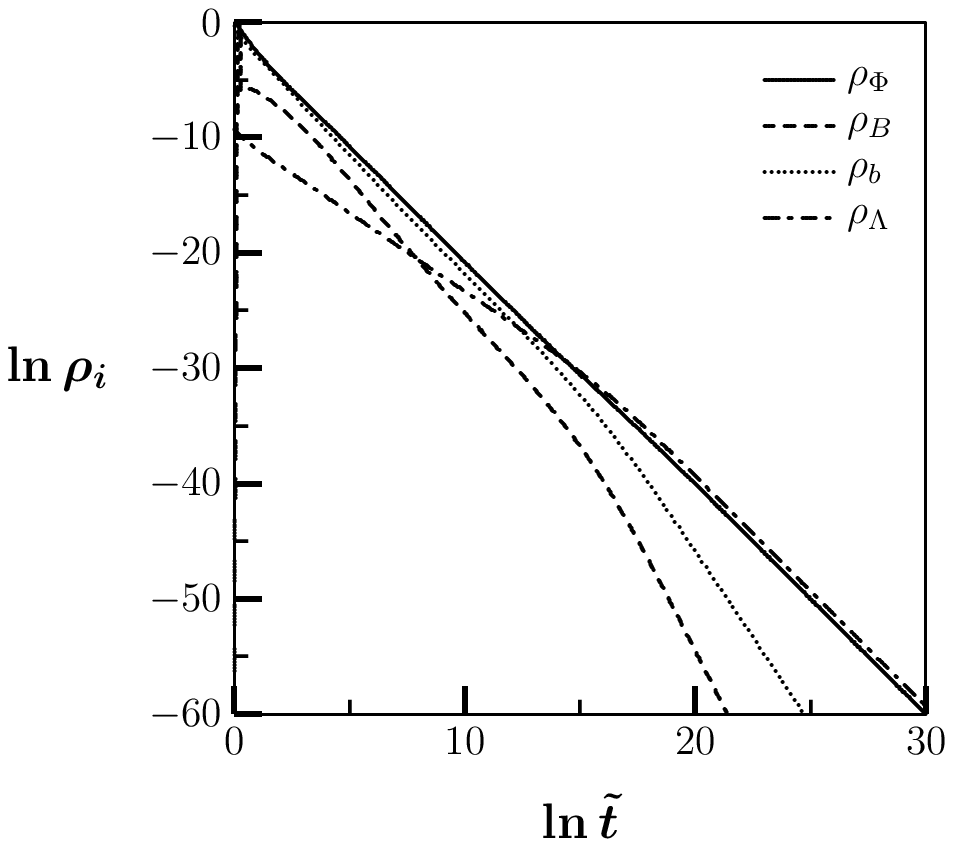}
\end{center}
\caption{Numerical solutions for $B\ne0$ and $\Lambda>0$.
When $\lambda=10^{-4}$, the solutions of four initial conditions are given:
$\dot\Phi_0=-7/4\cdot5^{1/4}$, $b_0=1/2$, $\dot B_0=0$ for the solid curves,
$\dot\Phi_0=0$, $b_0=1$, $\dot B_0=0$ for the dotted curves,
$\dot\Phi_0=0$, $b_0=10$, $\dot B_0=0$ for the dashed curves, and
$\dot\Phi_0=-\sqrt{3/2}$, $b_0=1$, $\dot B_0=0$ for the dash-dotted curves.
The lower-right panel shows the evolution of each component of energy density
for initial conditions $\dot\Phi_0=-\sqrt{3/2}$, $b_0=1$, $\dot B_0=0$.}
\label{fig7}
\end{figure*}

For $\Lambda>0$, the evolution is divided into two stages.
In the first stage where $\rho_b$ is dominant, the solution
approaches an attractor (\ref{sol-2}) of the $\Lambda=0$ case.
The approached value of $b$ is either $+\frac12$ or $-\frac12$
depending on the initial conditions.
If we look at the evolution of each component of
energy density, the kinetic energy of the dilaton $\rho_\Phi$ catches
up the potential energy $\rho_b$ and the ratio of them becomes constant.
This is a characteristic feature of the scaling solution \cite{Copeland:2006wr}.
The kinetic energy of $B$ field is kept much smaller than both of them,
but the anisotropy is still maintained due to the difference between $\rho_b$
and $\tilde\rho_b$.
In the second stage where $\rho_\Lambda$ is dominant,
it approaches another attractor (\ref{sol-3}).
Thus the universe recovers the isotropy.
The final value of $b$ is a certain constant which is determined by
initial conditions and can differ from $\pm\frac12$.
Numerical solutions for a few initial conditions
are shown in Figure~\ref{fig7}.
The kinetic energy of the dilaton $\rho_\Phi$ catches
up the potential energy $\rho_b$ in the first stage
and $\rho_\Lambda$ in the second stage.
The ratios, $\rho_\Phi/\rho_b$ and $\rho_\Phi/\rho_\Lambda$, approach
constants in each stage. The kinetic energy of $B$ field
is kept much smaller as in $\Lambda=0$ case.

\section{The Dilaton Stabilization}

\begin{figure*}
\begin{center}
\begin{picture}(500,430)(0,0)
\put(  0,240){\includegraphics[height=65mm]{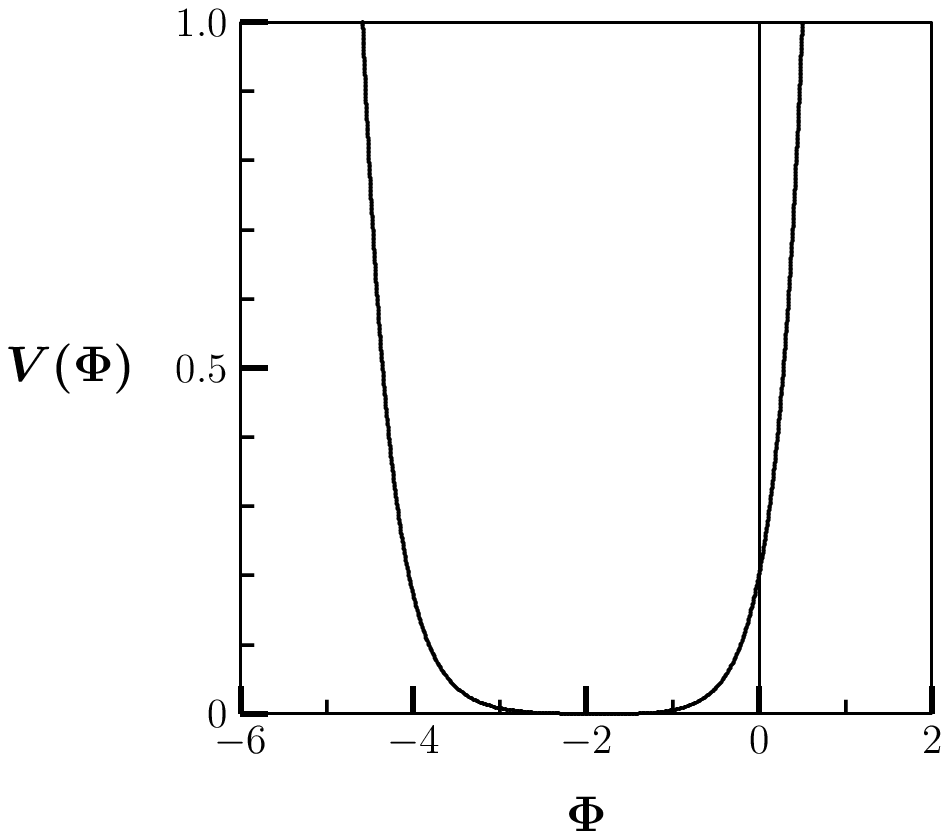}}
\put(248,240){\includegraphics[height=65mm]{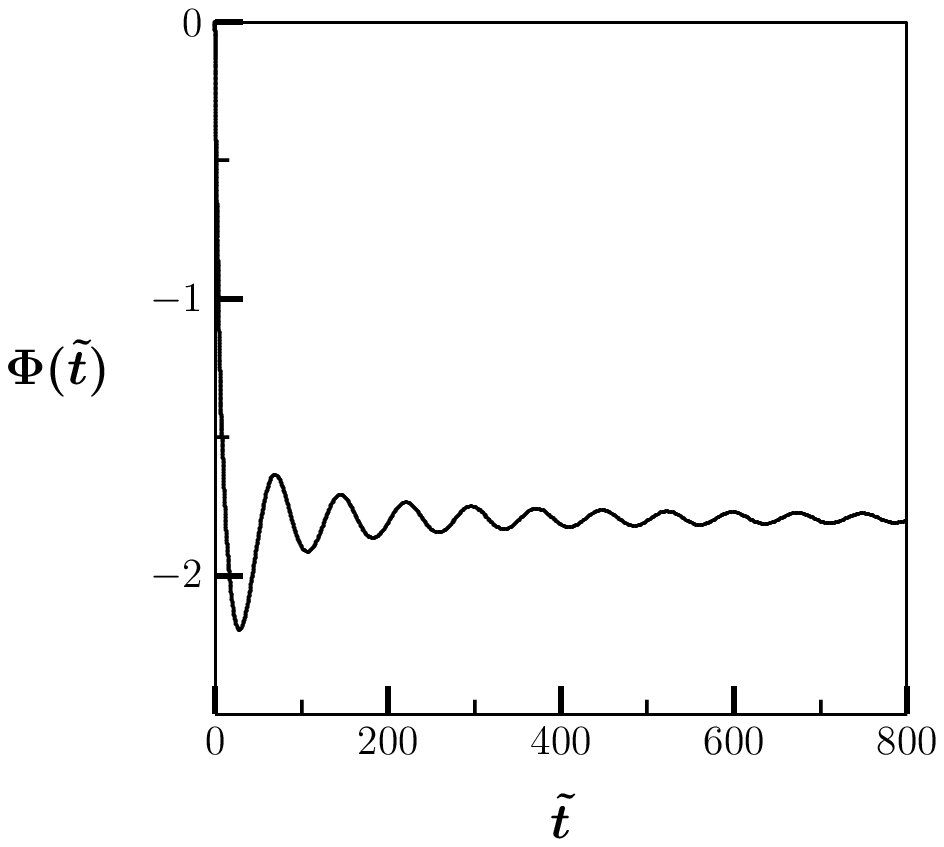}}
\put(  2,  0){\includegraphics[height=65mm]{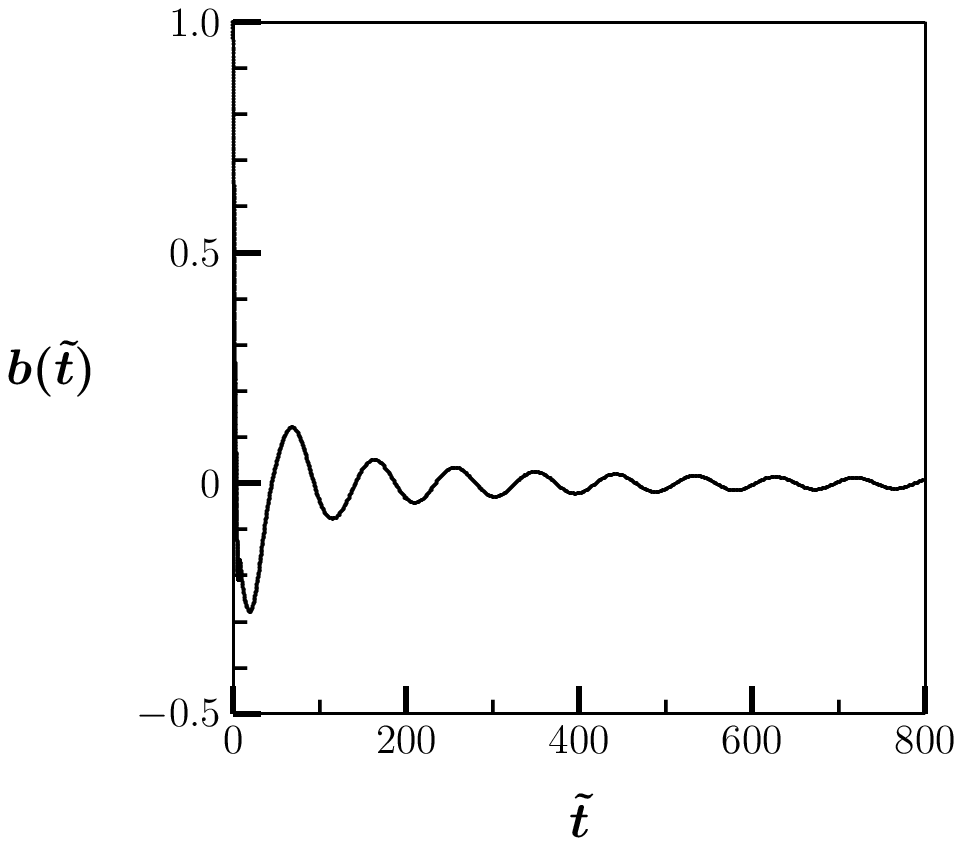}}
\put(240,  0){\includegraphics[height=65mm]{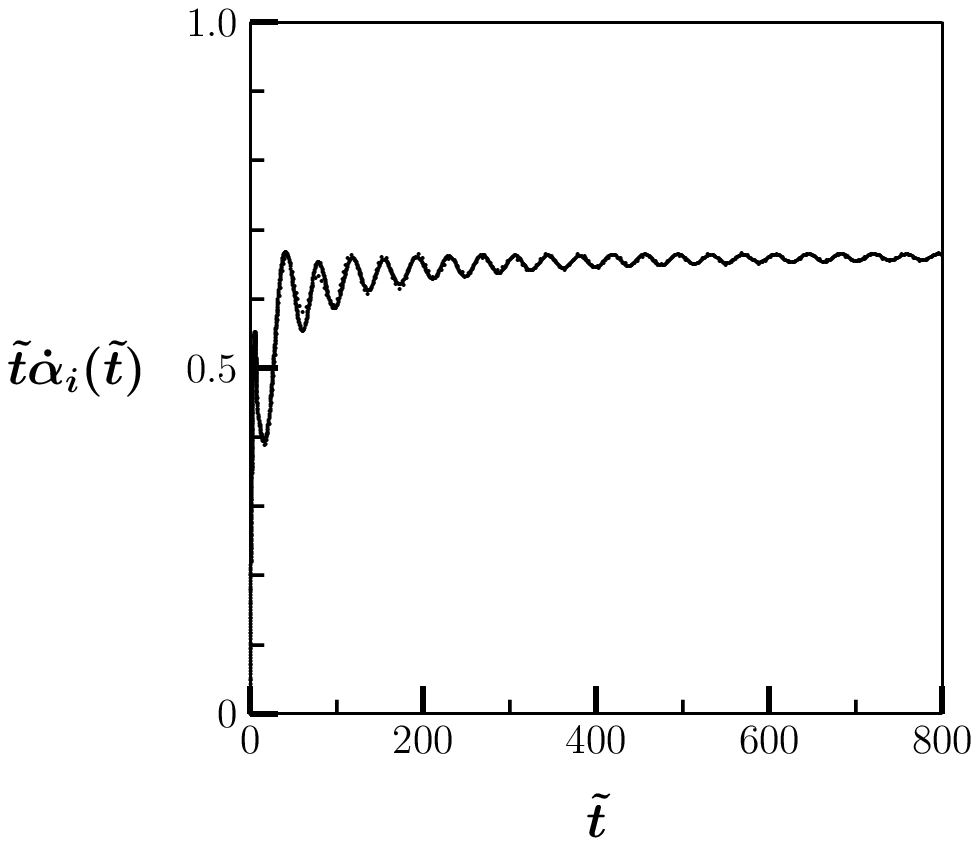}}
\put(120,435){(a)} \put(360,435){(b)} \put(120,195){(c)}
\put(360,195){(d)}
\end{picture}
\end{center}
\caption{(a) The potential $V(\Phi)$ for $\lambda=-0.1$.
The minimum of the potential is at $\Phi_{\rm m}=-\ln 6$ with $V(\Phi_{\rm m})=0$.
(b), (c) and (d) show the numerical solutions of $\Phi(\tilde t)$,
$b(\tilde t)$ and $\dot\alpha_i(\tilde t)$, respectively,
for $\lambda=-0.1$ and initial conditions $\Phi_0=0$, $\dot\Phi_0=0$
and $b_0=1$.
$\Phi(\tilde t)$, $b(\tilde t)$, and $\tilde t\dot\alpha_i(\tilde t)$
approach $\Phi_{\rm m}$, $0$, and $2/3$, respectively.}
\label{fig8}
\end{figure*}

The vacuum expectation value of the dilaton determines both the gauge
and gravitational coupling constants of the low energy effective theory.
Therefore, the dilaton must be stabilized at some stage of the evolution
for the action (\ref{action-E}) to have something to do with the reality.
In this section, we study the cosmological evolution
when the dilaton is stabilized.
As for the correct mechanism of dilaton stabilization,
the consensus has not been made yet.
Our goal here is to illustrate an example of the dilaton stabilization
and look into the effect of it on the dynamics of the NS-NS 2-form field
and the cosmological evolution, since the overall features of which are
insensitive to the detailed mechanism of stabilization.

Our starting point is the dilaton potential (\ref{dilaton-potential}).
This potential possesses the minimum for $\Lambda<0$,
but the value of the potential at the minimum is negative
and needs to be set to zero by fine-tuning of the constant shift.
However, the constant shift of the potential has no motivation
in the context of string theory.
Instead, we introduce a term $\frac14m_{{\rm F}}^2 e^{-3\Phi}$
in the potential, which can arise from the effect of various form field fluxes
in extra dimensions~\cite{Lukas:1996zq}.
Thus, our potential for the dilaton for $B=0$ looks like
\begin{equation}
V_{{\rm F}}(\Phi) = \frac14\left(m_B^2e^{3\Phi}+2\Lambda e^{2\Phi}
+m_{{\rm F}}^2e^{-3\Phi}\right).
\label{eq=V1}
\end{equation}
This potential has a global minimum for any value of $\Lambda$ and
$m_{{\rm F}}$.
To have sensible cosmology, the potential at the minimum must be zero.
For $\Lambda<0$, this can be done through a fine-tuning of the parameters
in the potential
\begin{equation}
\mu^2=\frac15\left(-\frac53\lambda\right)^6,
\end{equation}
where $\mu^2=m_{{\rm F}}^2/m_B^2$.
Then the minimum is located at
\begin{equation}
\Phi_{{\rm F}}=\ln\left(-\frac53\lambda\right).
\end{equation}
The shape of this fine-tuned potential for $\lambda=-0.1$ is shown
in Figure~\ref{fig8}-(a).
Now the equations (\ref{eq-phi})--(\ref{eq-a3}) are modified accordingly
\begin{equation}
\ddot\Phi + (\dot\alpha_1+\dot\alpha_2+\dot\alpha_3)\dot\Phi
= -2\rho_B-\rho_\Lambda-\frac12\rho_b-\tilde\rho_b+\frac32\rho_{{\rm F}},
\label{eq-phi-mod}
\end{equation}
\begin{eqnarray}
\ddot\alpha_1+\dot\alpha_1(\dot\alpha_1+\dot\alpha_2+\dot\alpha_3)
&=& \rho_\Lambda+\rho_b+\rho_{{\rm F}}, \\
\ddot\alpha_3+\dot\alpha_3(\dot\alpha_1+\dot\alpha_2+\dot\alpha_3)
&=& 2\rho_B+\rho_\Lambda+\tilde\rho_b+\rho_{{\rm F}},
\label{eq-a3-mod}
\end{eqnarray}
where $\rho_{{\rm F}}=\frac12\mu^2e^{-3\Phi}$,
while the equation (\ref{eq-b}) for $b$ is not changed.

With the stabilizing potential for the dilaton, the cosmological evolution
is completely changed. The dilaton and the NS-NS 2-form field
rapidly come to the oscillation about the potential minimum
$\Phi=\Phi_{{\rm F}}$ and $b=0$.
Oscillating $\Phi$ and $B$ fields behave like ordinary matter satisfying
the equation of state $p=0$ \cite{Chun:2005ee}.
The universe becomes isotropic and matter dominated.
The numerical solutions for the stabilizing dilaton potential (\ref{eq=V1})
with $\lambda=-0.1$ are plotted in Figure~\ref{fig8}-(b) to (d).
One can see the oscillation of $\Phi$ and $b$, and that both $\dot\alpha_1$
and $\dot\alpha_3$ approach to $2/3t$, indicating the matter domination.
Damping of $\Phi$ and $b$ oscillations is due to the expansion of the universe.

\section{Conclusion}

We investigated  cosmology of a four-dimensional low energy
effective theory arising from the NS-NS sector of string theory
with a D-brane which contains the dynamical degrees of freedom
such as the gravity, the dilaton, and the antisymmetric tensor
field of second rank, coupling to the gauge field strength living
on the brane.
The dilaton gains a potential in the presence of the D-brane,
the fluxes of various form fields and the curvature of extra dimensions.
The NS-NS 2-form field becomes massive in the presence of the D-brane.
The dynamics of the system crucially depends on the curvature parameter
$\Lambda$ and the brane tension parameter $m_B$
through which the dilaton obtains
the potential of the form; $\Lambda e^{2\Phi}+{1\over2} m_B^2 e^{3\Phi}$.
Here, the latter becomes the effective mass of the 2-form field.

When the 12-component $B(t)$ of the 2-form field is turned on,
the universe undergoes an anisotropic expansion
described by the Bianchi type-I cosmology.
We found the attractor solutions showing the overall features of general
solutions and confirmed it through numerical analysis.
The dilaton $\Phi(t)$ runs to the negative infinity
settling to a logarithmic decrease in time.
When the brane tension term $\frac12m_B^2 e^{3\Phi}$ dominates,
the anisotropy due to the 2-form field flux is sustained.
If there is a positive curvature term $\Lambda e^{2\Phi}$,
it dominates finally over the brane tension term
as the dilaton rolls down to the negative infinity.
Then the expansion of the universe turns to be isotropic and linear in time.
Accordingly, $B(t)$ decreases inversely proportional to time.

For sensible phenomenology and cosmology, the dilaton must be stabilized.
In order to study the dynamics of the 2-form field (B-matter)
and the stabilized dilaton system,
we adopted a dilaton mass term of the form $m_{{\rm F}}^2 e^{-3\Phi}$.
Then the dilaton potential has a global minimum
and the cosmological constant of our Universe can
be fine-tuned to a desired value with negative $\Lambda$.
With this stabilizing potential, we obtain a reasonable cosmology
from an initially anisotropic universe resulting from the 2-form field flux.
While the dilaton evolved to a stabilized value,
$B(t)$ shows an oscillatory matter-like behavior (B-matter),
and the universe expands as in the usual matter-dominated era
recovering the isotropy.

Finally, there are other dynamical degrees of freedom in the D-brane world,
such as the R-R form fields, which are not included in the present work.
Investigating the cosmological implications of them is left for the future work.

\begin{acknowledgements}
This work is the result of research activities
(Astrophysical Research Center for the Structure and
Evolution of the Cosmos (ARCSEC)) and was supported by grant
No.~R01-2006-000-10965-0 from the Basic Research Program
of the Korea Science \& Engineering Foundation (I.C. $\&$ Y.K.),
by the Science Research Center Program of
the Korea Science and Engineering Foundation through
the Center for Quantum Spacetime(CQUeST) of
Sogang University with grant number R11--2005--021 and
the grant No.~R01-2004-000-10520-0 from
the Basic Research Program of the Korean Science \& Engineering
Foundation (H.B.K.).
\end{acknowledgements}

\end{document}